\documentclass[iop]{emulateapj}

\newcommand{\myemail}{Duncan.Galloway@monash.edu}
\newcommand{\xte}{{\it RXTE}}

\newcommand{\src}{Sco~X-1}
\newcommand{\actaa}{AcA}

\newcommand{\nphot}{567}

\slugcomment{Accepted by ApJ}

\shorttitle{Precision ephemerides: I. Sco X-1}
\shortauthors{Galloway et al.}

\begin{document}

\title{Precision ephemerides for gravitational-wave searches: I. Sco
X-1\footnote{B\lowercase{ased on observations made with} ESO
\lowercase{Telescopes at the} L\lowercase{a}
S\lowercase{illa} P\lowercase{aranal} O\lowercase{bservatory under
programme} ID 087.D-0278.}}

\author{Duncan K. Galloway\altaffilmark{1,2,3,4}, 
  Sammanani Premachandra\altaffilmark{2},
   Danny Steeghs\altaffilmark{5},
   Tom Marsh\altaffilmark{5}, 
   Jorge Casares\altaffilmark{6,7},\\  and
   R\'emon Cornelisse\altaffilmark{6,7}}
\affil{\altaffilmark{1}Monash Centre for Astrophysics, Monash
  University, VIC 3800, Australia;
\altaffilmark{5}Department of Physics, Astronomy and Astrophysics group, 
University of Warwick, CV4 7AL, Coventry, UK;
\altaffilmark{6}Instituto de Astrof\'isica, E-38205, La Laguna, Tenerife,
  Spain}

\email{\myemail}

\altaffiltext{2}{also School of Physics, Monash University, VIC 3800, Australia}
\altaffiltext{3}{also School of Mathematical Sciences, Monash University, VIC 3800, Australia}
\altaffiltext{4}{ARC Future Fellow}
\altaffiltext{7}{also Departamento de Astrofisica, Universidad de La Laguna,
  E-38205, La Laguna, Tenerife, Spain}

\begin{abstract}
Rapidly-rotating neutron stars are the only candidates for persistent
high-frequency gravitational wave emission, for which a targeted search can be
performed based on the spin period measured from electromagnetic (e.g.
radio and X-ray) observations. 
The principal factor determining the sensitivity of such
searches is the measurement precision 
of the physical
parameters of the system. Neutron stars in X-ray binaries 
present additional computational demands for searches due to the
uncertainty in the binary parameters.
We present the results of a pilot study with the goal of improving the
measurement precision of binary orbital parameters for candidate
gravitational wave sources.  We observed the optical counterpart of Sco
X-1 in 2011 June with the William Herschel Telescope,
and also made use of Very Large Telescope observations in 2011, to provide
an additional epoch of radial-velocity measurements to earlier
measurements in 1999.
From a circular orbit fit to the combined 
dataset, we obtained an improvement of a factor of two in the
orbital period precision, and a factor of 2.5 in the epoch of
inferior conjunction $T_0$. While the new orbital period is consistent with the
previous value of \cite{gottlieb75}, the 
new $T_0$
(and the amplitude of variation of the Bowen line velocities) exhibited a
significant shift, which we attribute to 
variations in the emission geometry with epoch.
We propagate the uncertainties on these parameters through to the expected
Advanced LIGO \& VIRGO detector network observation epochs, and quantify
the improvement obtained with additional optical observations.
\end{abstract}


\section{Introduction}

Neutron stars 
orbiting low-mass stellar companions typically rotate many
hundreds of times every second \cite[e.g.][]{chak03a}, because the mass transferred within these
low-mass X-ray binary (LMXB) systems over their long ($\sim10^9$ year) lifetimes causes
the neutron star to spin up.
These extreme objects
must be highly spherical due to the
intense gravitational field; the equivalent of the tallest mountain
possible on Earth might be only a few centimetres high on a neutron star.
However, there are number of physical
processes thought to permit slight ($\sim10^{-6}$) 
deviations from axisymmetry,
for example from a temperature asymmetry arising from a non-aligned magnetic
field \cite[]{bil98c}. A quadrupole mass moment will lead to gravitational
wave emission at twice the neutron star spin frequency, $\nu_s$. 
Assuming a balance between spin-up torques and
angular momentum losses from gravitational wave emission, the expected gravitational wave
strength at the Earth from a distant source is proportional to the
accretion rate
(measured locally as the incident X-ray flux, $F_X$) and the spin frequency:
\begin{eqnarray}
h_0 & \approx & 4\times10^{-27}\, 
  \left(\frac{F_X}{10^{-8}\ {\rm erg\,cm^{-2}\,s^{-1}}}\right)^{1/2}\, 
\nonumber
  \\ && \times\left(\frac{300\ {\rm Hz}}{\nu_s}\right)^{1/2}
\end{eqnarray}
where 
$h_0=\Delta L/L$ is the ``strain'', i.e. the fractional change in length
of (for example)  an interferometer arm, 
and $\nu_s$ is the neutron star spin frequency.
The best targets for
searches for GW then, are the brightest sources with the lowest spin
frequencies. Unfortunately, the brightest neutron-star binaries,
the so-called ``Z-sources'' \cite[after their characteristic X-ray spectral
behaviour;][]{hvdk89},
 are also
those for which the spin periods are unknown. Spin measurements in about
20\% of known LMXBs have been made
by detecting various types of transient X-ray intensity pulsations
\cite[e.g.][]{watts12a};
however, none of these phenomena have yet been detected in the brightest class
of sources.

The principal difficulty in searching for the gravitational waves emitted by
neutron stars is the lack of precise knowledge about the neutron star spin.
This problem is compounded for sources in LMXBs
where one 
must also correct for the position and velocity of the neutron star in its
binary orbit
\cite[]{watts08}.
Without precise knowledge of the spin frequency and orbital parameters, a
search may still be carried out, but an observational ``penalty'' must be
paid. Effectively, the signal must be proportionately stronger, compared to a
source where the parameters are known more precisely, to reach the same
level of confidence for a detection. 

Contemporary (published) searches for the periodic gravitational wave emission
of these objects have adopted one of two techniques.  The first is a ``matched
filtering'' approach which involves comparing the observed signal from the
interferometer with a model signal coherently over some time interval
$T_{\rm obs}$ \cite[e.g.][]{ligobband07}.
This is currently the 
most sensitive method known (in the limit of
infinite computational power) and therefore the method of choice when not
computationally bound by a prohibitively large parameter space. For LMXB
sources, generally there is some large degree of uncertainty in the model
parameters, demanding the requirement for multiple models (``templates'' or
``filters'' in the GW search parlance). The second method is an example of a
semi-coherent approach, in this case involving the cross-correlation of the
outputs of two or more detectors coherently over short intervals, the products
of which are combined incoherently over the length of the observation
\cite[]{dhurandhar08,chung11}. Other
variations on the semi-coherent theme are in development at present but as is
the case for all of them, sensitivity is sacrificed for computational
feasibility and therefore the reason such approaches are used is entirely due
to the large uncertainties in the system parameters. 

Sco X-1, the brightest of the known LMXBs,
has already been the subject of two searches with data from the
initial LIGO detectors. In the first of these analyses~\cite[]{ligobband07}, a fully
coherent search using data from the 2-mo second science run (S2), 
the uncertainties in the orbital parameters and the unknown spin period
principally determined the search sensitivity, via the maximum duration (6~hr) of data that
could be searched coherently with the available computer power. 
This is a tiny fraction of the available observing time.
The second
analysis, a semi-coherent radiometer method, is a novel non-optimal approach
applied to Sco X-1 at a significant cost in sensitivity because of the large
uncertainty in the source parameters \cite[]{ligolimit07}. 

It follows that if the
system parameters were known precisely enough that the deviation between the
model and the target signal is small (less than one cycle) over $T_{\rm obs}$, a
fully coherent matched filter approach can be used and optimal sensitivity can
be achieved.

Sco X-1 is 
also one of the rare cases among the low-mass X-ray binaries where the
system is relatively unobscured and is bright in the optical band
($V\approx12.5$). The orbital period
is well known from 
89 years of photometry
\cite[]{gottlieb75}. Previous high-resolution spectroscopic studies of
Sco~X-1 led to the discovery of the first tracer of the unseen mass donor
star (\citealt{steeghs02}; hereafter SC02). These studies revealed
emission components from the irradiated low mass donor star,
principally within the Bowen fluorescence lines
near 4640~\AA{}. 

Despite the relatively high precision of the orbital parameters for Sco X-1, 
the current ephemeris has been derived from data obtained in 1999 or
earlier, and
thus will not have adequate precision to guide gravitational wave
searches in the A-LIGO era.
In order to
cover the possible parameter space in the search, the orbital parameter
uncertainties {\em at the time of the gravitational wave observation} must be
known. This can be estimated from previous measurements, given sufficient
information, such as the covariance matrix for the fit by which the
orbital parameters are measured; but this information is not currently
available.

Here we present 
initial
results from a
pilot program of optical observations of the stellar counterparts to
X-ray bright accreting neutron stars, in order to improve the precision
of the binary parameters.
These measurements will allow sensitivity
improvements for future gravitational wave searches, and will
also facilitate pulsation searches with the extensive X-ray timing
data from NASA's {\it Rossi X-ray Timing Explorer}\/ ({\it RXTE}), to
measure the spin frequency.

\section{Observations}

We summarise the source data for this paper in Table \ref{obsdata}.
We observed Sco X-1 for a second epoch on the nights of 2011 June 16--18,
using
the Intermediate dispersion Spectrograph and Imaging System (ISIS), at the
Cassegrain focus of the 4.2m William Herchel Telescope (WHT), La Palma.
ISIS is
a high-efficiency double beam spectrograph, providing medium
resolution spectra with dispersion in the range 8--120~\AA{}/mm.
For the new observations, we
adopted the H2400B grating on the blue arm in order to achieve a spectral
resolution of 0.32~\AA\ imaged with the $2048 \times 4096$ pixel EEV CCD detector.
We used a 1.0'' slit and obtained spectra covering a wavelength range of
4400--5000~\AA\  
at 0.11\ \AA\ per pixel.
At the central wavelength of 4700\ \AA, 
a resolution element spans
$20\ {\rm km\,s^{-1}}$.
A total of 157
spectra were obtained in 300-s exposures over three consecutive nights, covering
75\% of the 18.9~hr orbital cycle.

We reanalysed the spectra reported by SC02, also obtained with
ISIS on the WHT.
The R1200B grating was used for those observations, covering the
wavelength range 4150--5050~\AA\ at 0.45~\AA\ per pixel. The spectra
were recorded using a $2048\times4096$~pixel EEV CCD detector; a
$1\farcs2$ slit gave a spectral resolution of 0.84~\AA, corresponding to
$55\ {\rm km\,s^{-1}}$ at the central wavelength of 4601~\AA.
A total of 137 spectra were obtained over the nights of 1999 June 28--30,
covering 75\% of the orbital cycle. For full details of those
observations, refer to SC02

We further observed Sco~X-1 in a series of observations with the
Ultraviolet and Visual Echelle Spectrograph (UVES), on the
European Southern Observatory (ESO) Very Large Telescope (VLT)
between 29 May--23 August 2011 (programme 087.D-0278).
UVES is a two-arm
cross dispersed echelle spectrograph, providing a resolving power up to
80,000 in the blue arm 
\cite[]{dekker00}.
A total of 44 exposures lasting 720~s each were made
using a $2048\times4096$~pixel EEV  CCD detector,
covering the wavelength range 3000--5000~\AA\  at  0.0294~\AA/pixel. A
1~arcsec slit was used, giving spectral resolution of 
0.1~\AA\
(8~${\rm km\,s^{-1}}$).

The WHT spectral data were reduced using a series of
{\sc pamela}\footnote{ {\url
http://deneb.astro.warwick.ac.uk/phsaap/software} } routines.
Raw frames were first debiased using a median bias frame calculated from
combining 20--25 individual bias exposures.
Flat field correction was achieved using median of 
$\approx40$ 
Tungsten exposures constructed each night. The frames were then
flat fielded, this corrected for pixel-to-pixel variations in responsivity
in order to normalize the background counts and help remove systematically
hot or dead pixels.
CuNe and CuAr arc images were observed 
for the purpose of wavelength calibration.  
Arc images were taken regularly throughout each night, with
wavelengths solutions for each source spectrum interpolated from the
nearest arcs before and after.
The long slit was rotated
to accommodate a nearby comparison star in order to monitor
slit-losses.

Sky subtraction
was achieved by selecting
the sky and object regions and fitting polynomials in order to estimate
the sky under the object. 
The target and comparison star spectra were
then optimally extracted for each exposure.
Finally, a wide slit exposure of the flux standards BD284211 and MHZ44
\cite[]{oke90} was used to perform flux calibration using {\sc
molly}$^{1}$ data
analysis package.

UVES echelle spectra were reduced using the pipeline provided by
ESO\footnote{ \url{http://www.eso.org/sci/software/pipelines}}. After
reaching the last stage of data reduction, we transformed the spectra
using IRAF into a format that can be read by the {\sc molly} software package. 

Following SC02, we analysed the narrow emission lines arising from
individual Bowen transitions, which are superimposed on a broader
emission component centred around 4640~\AA.
The strongest Bowen components are the C III component at 4647.4 \AA{} 
and the two N III components at 4634.1 
and 4640.6~\AA{}.
These Bowen components move in phase with each other and
move in antiphase with the He II emission (SC02).
 
All the individual 
338
spectra were normalized to the continuum. Following
SC02, we fitted
a model consisting of a broad underlying Gaussian component and three
narrow components, with 
central wavelengths for the lines having a common offset corresponding to
the line-of-sight radial velocity.
We fixed the widths of the line components to values 
used in SC02;
the full-width at half maximum (FWHM) was fixed to
$1250\ {\rm km\,s^{-1} }$
for the broad component and 
$75\ {\rm km\,s^{-1} }$
for the narrow components.
The 
model normalisation for 
each line was free to vary independently.

We also used $V$-magnitude aperture photometry of \src\ made for the All-Sky
Automated Survey \cite[ASAS;][]{pojmanski04}.
This program provides continuous monitoring of the whole sky with
automated instruments located in Chile and Hawaii. Each
instrument consists of a wide-field ($9^\circ\times9^\circ$) camera equipped
with an f200/2.8 telephoto lens; photometric magnitudes are determined with
a range of apertures (2--6 pixels in diameter) for all stars to a limiting
magnitude of $V\approx14$.
We obtained a total of 640 measurements of \src\ between 2001 and 2009.
The mean uncertainty was 0.047 magnitudes.

\begin{deluxetable}{lccl}
\tablecaption{Observations of Sco X-1, 1999--2011
 \label{obsdata}}
\tablewidth{0pt}
\tablehead{
  \colhead{Instrument/source}
  & \colhead{Date}
  & \colhead{$n_{\rm obs}$}
  & \colhead{Ref.}
}
\startdata
ISIS/WHT & 1999 Jun 28--30 & 137 & [1] \\
ASAS\tablenotemark{a} & 2001 Jan 22--2009 Oct 5 & \nphot\tablenotemark{b} & [2]\\
UVES/VLT & 2011 May 29--Aug 23 & 44 & [3] \\
ISIS/WHT & 2011 Jul 16--18 & 157	& [3] 
\enddata
 \tablerefs{ 1. SC02, 2. \cite{pojmanski04}, 3. this paper }
 \tablenotetext{a}{ {\url http://www.astrouw.edu.pl/asas} }
 \tablenotetext{b}{ Grade A \& B measurements, retained from 640 total
measurements }
\end{deluxetable}

\section{Analysis and results}
\label{sec1}

 \subsection{Photometric variations}
 \label{subsec1}

\begin{figure}
 \plotone{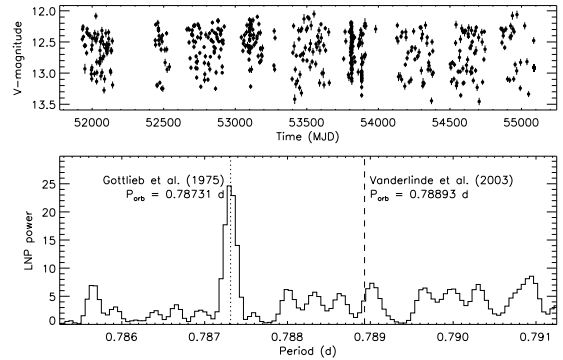}
 \figcaption[]{$V$-band photometry of \src\ from the All-Sky Automated
Survey (ASAS). The top panel shows the raw data, consisting of \nphot\ 
grade A \& B measurements between 2001 and 2009.
The bottom panel shows the Lomb-normalised periodogram of the data. The
candidate orbital period is shown as the dotted line \cite[]{gottlieb75}
and the alias from the \xte/ASM data is shown as the dashed line
\cite[]{vanderlinde03}. Note the excess of power at 0.787313~d; the
estimated significance is $1.7\times10^{-7}$, equivalent to $5.1\sigma$.
 \label{photometry} }
\end{figure}

In order to identify the correct orbital period from the candidates of
\cite{gottlieb75} and \cite{vanderlinde03}, we first performed a periodicity
search of the $V$-band photometry from ASAS.
We plot the $V$-magnitude for the optical counterpart to \src\ in Fig.
\ref{photometry} ({\it top panel}). We adopted the magnitudes measured
with the smallest aperture (2~micron); our results were only
weakly sensitive to the choice of aperture. The mean intensity was
$V=12.7\pm0.3$. 

We calculated a Lomb-normalised periodogram in the frequency range
$10^{-6}$--1.34~d$^{-1}$. The mean power thus measured was 2.95, with
standard deviation 2.57. The maximum power in this range was 24.6, at a
period of 0.78730~d (Fig. \ref{photometry}, {\it bottom panel}), where the
resolution was approximately $5\times10^{-5}$~d. This period is in good
agreement with the value of \cite{gottlieb75}; in contrast, the maximum
power measured within 0.001~d of the alias value of 0.78893~d was 7.3.
This value is consistent at the $1.7\sigma$ level with the full-frequency
range distribution of powers.

Recently a similar analysis was reported adopting the same data
\cite[]{hynes12}. Our results are identical to that work. \cite{hynes12}
derived a period of $0.787313(15)$~d, which is consistent with the
\cite{gottlieb75} period \cite[but not the value of][]{vanderlinde03},
confirming the former value as the orbital period for \src. The epoch of
minimum light for the ASAS data is $T_{\rm min}({\rm HJD}) = 2453510.329(17)$,
consistent to within the uncertainties with the projected
\cite{gottlieb75} ephemeris.
Below we present additional analysis leading to further refinement of
these orbital parameters.

 \subsection{Bowen blend spectroscopy}
 \label{subsec2}

The earlier measurement of the orbital period of Sco~X-1
\cite[]{gottlieb75} was based on such a long baseline (84.8~yr) that we
did not initially anticipate being able to improve on the precision.
Our optical spectroscopic data were limited to a much shorter timespan,
just 12~yr. However, the precision of the orbital period measurement
also depends on how precisely the orbital phase can be measured at any
epoch. For the photometric data, the phasing is rather poor, due to the
low measurement precision and the intrinsic photometric variability. On
the other hand, the results of SC02 suggest that for a few days of radial
velocity measurements, the orbital phase can be measured to approximately
$\Delta T_0=0.003$~d, or 0.4\%. Assuming we could make an equally good
measurement from our new data, the achievable precision can be estimated
as $\Delta P_{\rm orb} \approx \sqrt{2(\Delta T_0)^2}/n_{\rm cyc}$, where
$n_{\rm cyc}$ is the number of orbital cycles between the two
measurements. 
Our WHT observations in 2011 June were approximately 
5550~cycles	
after those of
1999~July, suggesting an achievable precision of $\Delta P_{\rm
orb}\approx 7\times10^{-7}$~d. This precision is 30\% better than 
that quoted by \cite{gottlieb75}, motivating a combined analysis of the
two datasets.

\begin{figure}
 \plotone{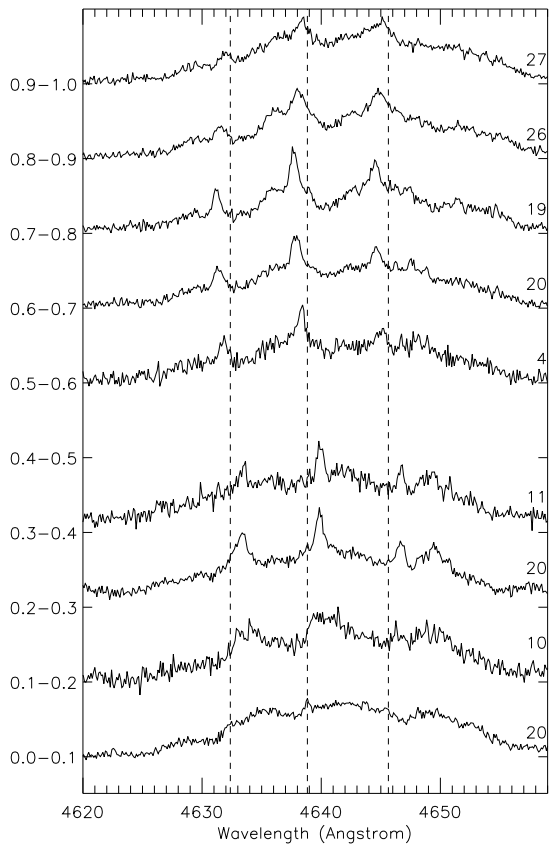}
 \epsscale{1.2}
 \figcaption[]{Averaged spectra from the 2011 June WHT observations of \src,
plotted as a function of orbital phase. Each plotted spectrum is the
average of all those observed in the phase range listed on the $y$-axis.
The number of individual spectra comprising each plotted spectrum is shown
against the right-hand $y$-axis. The spectra have been scaled and offset
vertically to illustrate the changing strength of the Bowen lines.
The dashed lines show the (system) rest-frame wavelengths for the Bowen
lines, shifted by the systemic velocity $\gamma$ (Table \ref{newparam}).
Note that no spectra were observed with WHT in the
phase range 0.4--0.5 in 2011.
 \label{spectra} }
\end{figure}

We first examined the WHT spectra to determine the properties of the Bowen
lines.
We show the spectrum within the wavelength range of interest, grouped and
averaged within orbital phase bins of width $\Delta \phi=0.1$
(according to the ephemeris of SC02),
in Figure \ref{spectra}. The lines are strong and clearly visible in the
phase range 0.1--0.9, but become weak around phase zero. At this phase,
we face
the unirradiated side of the secondary, also
corresponding to the phase of
minimum light.
It is thus not unexpected that the
strength of the emission lines is weakest at this phase compared to 
other phases. 

\begin{figure}
 \epsscale{1.1}
 \plotone{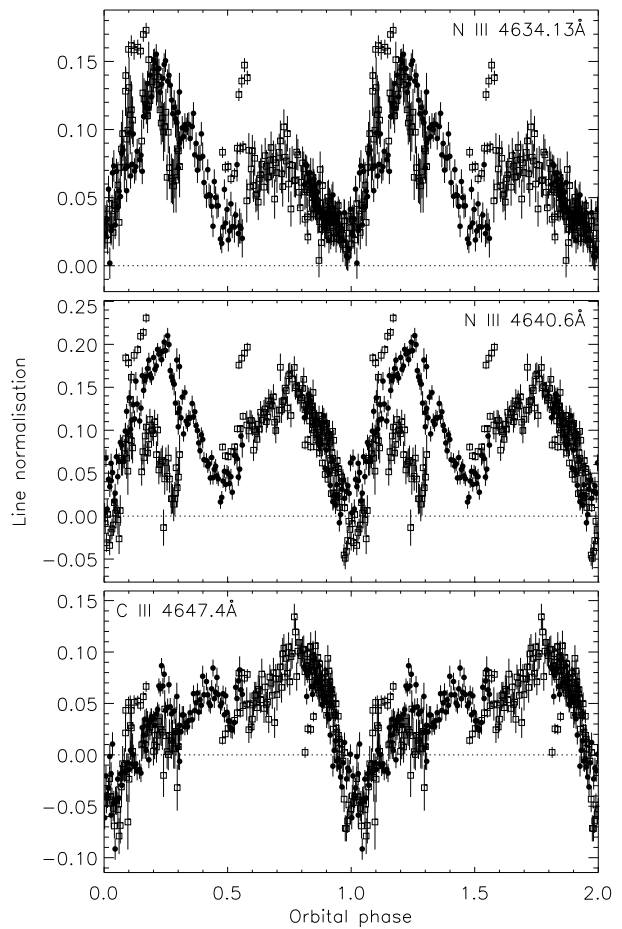}	
 \figcaption[]{Best-fit normalisation (relative to the continuum) for each
of the Bowen lines (4634.13\AA, 4640.6\AA\ and 4647.4\AA) in Sco X-1, from
the 1999 and 2011 WHT observation epochs.  The 1999 observations are shown
as filled circles, while the 2011 observations are shown as open squares.
Two cycles are shown in each panel for clarity.
Note the marked difference in the line strength both between and within
the two main observing epochs, particularly for the N{\sc III} 4640.6\
\AA\ line within phase ranges 0.1--0.3 and 0.5--0.6.
\label{linestrength} }
\end{figure}

We examined the fitted 
normalisation
for each of the lines individually in
Fig. \ref{linestrength}. The intensity of the 4634.13\AA\ and
4640.6\AA\ (N{\sc iii}) lines show a marked double-peaked variation, with
the maximum at phase 0.25, and primary and secondary minima at
phase 0.0 and 0.5, respectively.
In contrast, there was no secondary minimum for the
4647.4\AA\ (C{\sc iii}) line intensity, which was also the weakest of the
three. The best-fit normalisation
exhibited an asymmetry opposite in sense to the N{\sc iii} lines, reaching
a maximum at around phase 0.8.
These lightcurves differ substantially from 
the single peak at phase 0.5 expected for uniform illumination of the front
face of the  donor's Roche lobe.
It seems plausible that
some of the structure in our lightcurves may be caused by systematic
errors introduced by the (sometimes poorly constrained) broad component
underlying the Bowen lines.
However, inspection of the trailed spectra appears to confirm that the 
strong peak at phase 0.25 is real, as is the (weaker) peak at 0.75. Taking
the lightcurves at face value, it appears that 
the emission may not be symmetric around the $L_1$ point, with both sides
contributing, and the side away from the gas stream brightest.

\begin{figure}
 \plotone{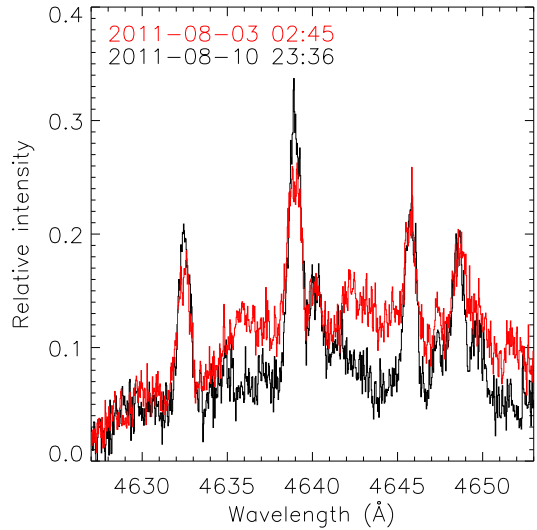}	
 \figcaption[]{Example VLT spectra showing significant variations in Bowen
line strength on timescales of a week. The spectra shown were observed by
the VLT in 2011, on Aug 3 (orbital phase $\phi=0.570$) and Aug 10
($\phi=0.565$). Note the dramatic difference in the strength of the lines
between the two spectra, despite the closeness in time and orbital phase.
 \label{epochdiff} }
\end{figure}

Further clouding the pattern of line 
intensity
variability was evidence
for epoch-to-epoch variations. For the 4640.6\AA\ (N{\sc iii}) line (Fig.
\ref{linestrength}, middle panel), the
relative normalisation measured by WHT around phase 0.2 in 2011 was significantly lower
than in 1999. All these data were taken on the same night, which might
suggest that a calibration issue present only on that night gave rise to
the lower 
intensities.
However, the best-fit normalisation for the other two lines
measured on that night are
more consistent between epochs. Furthermore, VLT data taken in an
overlapping phase range just one month later shows significantly {\it
higher} 
intensity
for this line. This is also seen between phase 0.5--0.6,
when VLT measurements taken at two epochs (in 2011 August, separated by
just 8 days) exhibit intensities that vary by approximately 55\%, far
in excess of the estimated uncertainty.
This is illustrated in the example spectra plotted in Fig. \ref{epochdiff}.
These variations are thus intrinsic to the source and suggest changes in
the emission line regions occur on non-orbital timescales.

\begin{figure}
 \plotone{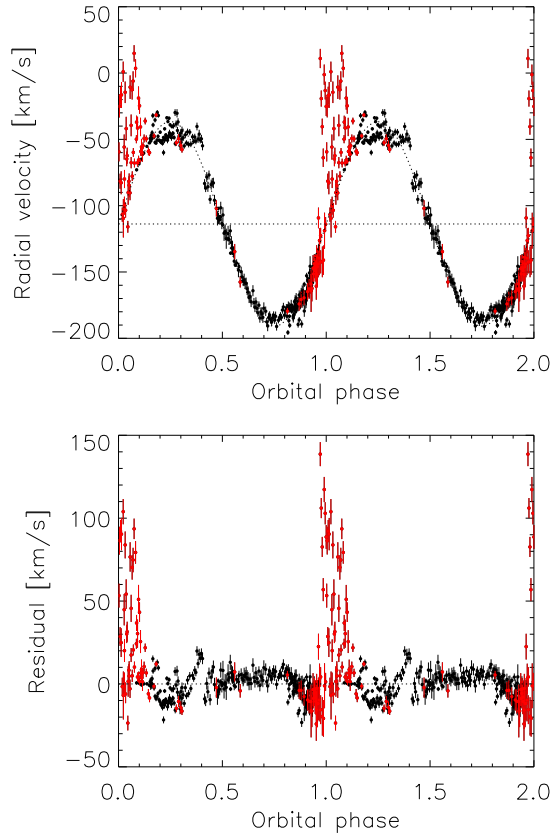}	
 \figcaption[]{Radial velocities for Sco X-1 determined from the 1999 and
2011 WHT data, and 2011 VLT measurements. In the upper panel we show the
radial velocities with the orbital model of SC02 overplotted as a dotted
sinusoid. The
systemic velocity is also shown as a horizontal dotted line. The points
are coloured according to the minimum significance of the line detection,
$S_{\rm min}$ (see text). Measurements with $S_{\rm min} \leq 2$ are shown in
red, while those with $S_{\rm min}>2$ are plotted as black. The data are
plotted twice for clarity.
The lower panel shows the residuals once the model is subtracted. Note the
very significant residuals around phase zero.
Error bars indicate the statistical 1-$\sigma$ uncertainty.
 \label{rvs} }
\end{figure}

The weakness of the lines 
around phase 0
had a dramatic effect on the radial velocity fits. We show in Figure
\ref{rvs} the fitted radial velocities for the combined 1999 and 2011
data.
The measurements deviate very significantly
from the predicted
values near the phase range 0.9--1.0 and 0.0--0.1. 
Within these phase ranges, the narrow
components were 
much weaker than at other phases,
leading to  
unreliable radial velocity measurements.
Similar deviations were present in the original analysis of the 1999 data
(see Fig. 3 of SC02),
although nowhere near as large. Here
the higher spectral resolution for the 2011 data may have led to a lower
signal-to-noise for the weak lines in any given bin.
However, we repeated our radial-velocity measurements with the 2011
spectra rebinned to the same resolution as the 1999 observations, and
found no reduction of the systematic errors. Thus, it may instead be the
lower 
line intensity measured in 2011 that played the main role in
determining the systematic errors.

We screened the data to
eliminate those unreliable velocities, by calculating the minimum
detection significance for  three lines. For each line $i$, we estimate
the significance as $A_i/\sigma_i$, where $A_i$ is the fitted line
normalisation and $\sigma_i$ the uncertainty (recall that the line width is
fixed). 
We calculated 
\begin{equation}
S_{\rm min} = {\rm min} ( A_1/\sigma_1, A_2/\sigma_2, A_3/\sigma_3)
\end{equation}
and rejected any measurements with $S_{\rm min}\leq 2$ (plotted in red in Fig.
\ref{rvs}). In this manner, we identified 
96
of the 338 WHT \& VLT measurements
as unreliable.

We fitted the remaining 
242	
measurements with a sinusoid, offset by the
systemic velocity, $\gamma$. 
We varied the input $T_0$ value to obtain the smallest possible cross-term
in the covariance matrix between $T_0$ and $P_{\rm orb}$ (see below).
We obtained a best-fit 
reduced-$\chi^2$ value of 
8.5 for 238	
degrees of freedom (DOF). The high $\chi^2$ indicates that
systematic errors are still likely present at a significant level, and so
we scaled the measurement errors by 
$\sqrt{8.5}$
in order to
obtain a reduced $\chi^2_\nu=1.0$ and estimate conservative parameter
uncertainties. We obtained the following system parameters
\begin{eqnarray}
\gamma & = & -113.8\pm0.5\ {\rm km\,s^{-1}}\nonumber \\
K & = & 74.9\pm0.5\ {\rm km\,s^{-1}}\nonumber \\
T_0 & = &  2454635.3683\pm0.0012\ ({\rm HJD})\nonumber \\	
P_{\rm orb} & = & 0.7873114\pm0.0000005\ {\rm d}\nonumber
\end{eqnarray}
where the uncertainties (here and throughout) are at the 1-$\sigma$ (68\%)
confidence level.
Rescaling the uncertainties in this way implicitly assumes that the
uncertainties are systematically underestimated. However, here there is
likely also a contribution of purely systematic errors, which should
instead be added in quadrature to the measurement uncertainties. To
estimate the upper limit of such a contribution, we introduced such a
systematic contribution and varied the magnitude until we again reached a
reduced $\chi^2$ of approximately 1. The required contribution was $6.2\
{\rm km\,s^{-1}}$; the resulting fit parameters were consistent with those
obtained by simply rescaling the errors, to within $\approx1\sigma$.

The precision of $P_{\rm orb}$ has been improved by
approximately 50\%, more than expected given our initial
predictions. This is likely due to the addition of the VLT data, which
offers slightly better radial velocity precision; the average
(statistical) uncertainty for the VLT measurements for which the line
significance was above the threshold was $2.2\ {\rm km\,s^{-1}}$, compared
to $3.2\ {\rm km\,s^{-1}}$ for WHT.
The new value is discrepant with that of
\cite{gottlieb75} at only the $1.3\sigma$ level.

The systemic velocity $\gamma$ is also consistent to within the uncertainties
with the previous value of SC02. However, the velocity
amplitude $K$ is significantly 
($4.4\sigma$)	
lower. 
We initially attributed this discrepancy to the screening of the radial
velocity measurements.  The systematic uncertainties contributed by the
weakness of the spectral
lines around phase zero tended to result in radial velocities
significantly higher than predictions. Together, these systematic errors
could be expected to bias the velocity amplitude to higher values (as well
as potentially shifting the $T_0$ earlier). 
To test this hypothesis, we performed radial velocity fits to the 1999
observations alone, with the orbital period fixed at the
\cite{gottlieb75} value (essentially replicating the analysis of SC02).
The velocity amplitude without (with) screening was
$77.2\pm0.4\ {\rm km\,s^{-1}}$
($76.4\pm0.3\ {\rm km\,s^{-1}}$). While the amplitude obtained using all
the 1999 measurements
was identical to the analysis of SC02, the value adopting the screening
criteria was discrepant at less than the $2\sigma$ level. Thus, it seems
likely that the discrepancy between the best-fit velocity amplitude for the
combined data and the results of SC02 is instead related to a change in
the geometry of the line emission region between epochs.

To compare the newly-derived $T_0$ with that of SC02, we
subtracted an integer number of orbital cycles
(4162) to give fractional days 0.5783
compared to 0.568.
Again, the discrepancy is significant
($3.2\sigma$,
not taking into account the contribution of the projected
uncertainty on $P_{\rm orb}$) but the sense is in the direction that we
expect from the known radial velocity bias around phase zero. Thus, we
expect that our new $T_0$ is more reliable than that of SC02, although
systematic uncertainties arising from changes in the emission geometry
from epoch to epoch may yet contribute to this discrepancy.

Similarly, we subtracted 
1429
orbital cycles to compare with the ephemeris
of \cite{hynes12}. We obtain fractional days of 0.3003 compared to 0.329,
which is consistent (at the 
$1.7\sigma$
level). The relative phasing of
the photometric minimum $T_{\rm min}$ is $0.039\pm0.017$, which is
also consistent with the relative phasing of the \cite{gottlieb75} data as
calculated by SC02.

The best-fit orbital parameters are listed in Table \ref{newparam}, along
with the
epoch of inferior conjunction 
$T_0$ in units of GPS seconds\footnote{GPS time zero
is 1980 Jan 6 00:00 UT (JD~2444244.5), and since it is not perturbed
by leap seconds is now ahead of UTC by 16 seconds as of 2012 July 1
--- see {\url http://hpiers.obspm.fr/iers/bul/bulc/bulletinc.dat}}.
We also quote the off-diagonal term from the covariance matrix $V(P_{\rm
orb},T_0)$ giving the
cross-term between $P_{\rm orb}$ and $T_0$, critical for determining the
propagated error on the future epoch of inferior conjunction.
We caution that the measured velocity amplitude $K$ is likely an
underestimate of the velocity amplitude of the companion's centre of mass,
as the line emission is dominated by the heated face of the companion
\cite[e.g.][]{munoz-darias05}. We
also defer any attempt to refine the projected velocity of the neutron
star itself.

One additional systematic uncertainty that must be considered arises from the
possibility that the centre of light for the Bowen lines does not lie on
the line joining the centres of mass of the two stars. Such a situation
might arise from asymmetries in the accretion stream or impact point,
leading to differential illumination on the leading compared to the
trailing side of the donor. In that case, the true inferior conjuction
would occur earlier or later compared to the epoch inferred from the Bowen line
spectroscopy.
Since the distribution of emission across the donor is unknown,
and 
is likely variable on non-orbital timescales, it is not 
currently feasible to construct a detailed model.
Instead we derive here some limits on
the effect of such an emission asymmetry on the ephemeris.
Firstly, we note that the
system parameters most consistent with the spectroscopic measurements and
the inferred system inclination have mass ratio $q\approx0.3$ and
$K_2\approx120\ {\rm km\,s^{-1}}$. With the measured $K$-amplitude for the
Bowen lines much smaller, at $75\ {\rm km\,s^{-1}}$, the emission region
must be close to the $L_1$ point. The angular size of the region on the
donor with this velocity is approximately $8^\circ$. Secondly, the degree of
the asymmetry on the donor is restricted by the relative emission at
different phases. Since we see emission at both phase 0.25 and 0.75, the
line emission cannot arise from one side alone; with a typical contrast of
50\% between the line amplitudes at these phases, the maximum shift is
likely no more than half of the region's angular size, or $\pm2^\circ$. This
corresponds to a phase error of 0.005~d, or approximately 4 times the
(statistical) error attributed to $T_0$ (Table \ref{newparam}).

\begin{deluxetable}{lcl}
\tablecaption{Orbital parameters for \src\ derived from 
multi-epoch fits to
radial-velocity measurements
 \label{newparam}}
\tablewidth{0pt}
\tablehead{
  \colhead{Parameter}
  & \colhead{Value}
  & \colhead{Units.}
}
\startdata
$\gamma$      & $-113.8\pm0.5$          & ${\rm km\,s^{-1}}$ \\
$K$         & $74.9\pm0.5$            & ${\rm km\,s^{-1}}$ \\
$T_0$         & $ 2454635.3683\pm0.0012$   & HJD \\
              & $897771000\pm100$       & GPS seconds \\
              & 2008 June 17 at 20:50:00 & UTC \\
$P_{\rm orb}$ & $0.7873114\pm0.0000005$ & d \\
$|V(P_{\rm orb},T_0)|$ & $3.614\times10^{-11}$ & d$^2$ 
\enddata
\end{deluxetable}

\subsection{Parameters for future GW searches}
\label{future}

Here we establish the orbital parameters and uncertainties likely to apply
for future searches for gravitational waves.
The current scenario for the operation of the Advanced LIGO-Virgo (aLIGO) network over the
next decade involves a series of commissioning periods with increasing
sensitivity beginning in 2015, leading to the operation of the full
network with full sensitivity by 2019 \cite[]{ligoprospects13}.

We note that the orbital phase for $T_0$ corresponds to inferior
conjunction of the mass donor, that is, at $T_0$ the companion is closest
to Earth, and the compact object is at its most distant. If the quoted orbital
parameters are to be used for searches for gravitational waves, the
relative phasing should be taken into account, and an offset of $0.5P_{\rm
orb}$ added to the reference phase if necessary (depending on the
expression for the orbital motion).

\begin{figure}
\epsscale{1.2}
 \plotone{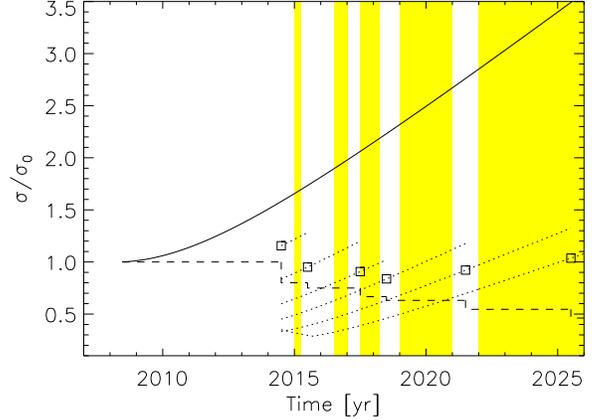}	
 \figcaption[]{The projected uncertainty in the epoch of inferior
conjunction $T_0$ ({\it solid line})
throughout the aLIGO
observing intervals ({\it shaded bands}). Note the approximately linear
growth of the effective error throughout the aLIGO commissioning and
observing period (2015--), in the absence of additional epochs of optical
observations. The approximate effect 
of additional observing epochs ({\it open squares})
is shown on the $T_0$ uncertainty ({\it dotted curves})
and the orbital period $P_{\rm orb}$ ({\it dashed line}).
 \label{errevolution} }
\end{figure}

In the absence of significant orbital period evolution, 
the orbital period uncertainty measured here is the correct value to use for
future searches. However, the correct $T_0$ uncertainty at a future epoch
will depend on the time elapsed since the measurement, as well as the
uncertainties on both $T_0$ and $P_{\rm orb}$. 
We here describe how the error on 
$T_0$ may be projected in time to any given epoch. The epoch for inferior
conjunction $T_n$ 
is given by
\begin{equation}
T_n = nP_{\rm orb} + T_0
\end{equation}
where $P_{\rm orb}$ and $T_{0}$ are given in Table \ref{newparam} and $n$
is an integer. 
The uncertainty $\sigma_n$ is thus
\begin{eqnarray}
\sigma_n^2 & = & \left(\frac{\partial T_n}{\partial P_{\rm
orb}}\right)^2\sigma_{P_{\rm orb}}^2  + \left(\frac{\partial T_n}{\partial
T_0}\right)^2\sigma_{T_0}^2 \nonumber \\ 
&&+ 2\left(\frac{\partial T_n}{\partial P_{\rm
orb}} \frac{\partial T_n}{\partial T_0}\right)V(P_{\rm orb},T_0) \\
           & = & n^2 \sigma_{P_{\rm orb}}^2 + \sigma_{T_0}^2 + 2nV(P_{\rm
orb},T_0)
\end{eqnarray}
where $\sigma_{P_{\rm orb}}$ and $\sigma_{T_0}$ are the uncertainties in
$P_{\rm orb}$ and $T_0$ respectively, and $V(P_{\rm orb},T_0)$ is the
cross-term of the covariance matrix (from Table \ref{newparam}).
Substituting in the values from Table \ref{newparam}, we express the error
in $T_n$ as 
\begin{eqnarray}
\sigma_{0,t} & \approx & [3.9\times10^{-13} (t/1\ {\rm d})^2 + 
       9.2\times10^{-11} (t/1\ {\rm d}) \nonumber \\ && +
       1.4\times10^{-6}]^{1/2}\ {\rm d} \\
   & = & [5.3\times10^{-8} (t/1\ {\rm yr})^2 + 
       3.4\times10^{-8} (t/1\ {\rm yr}) \nonumber \\ && +
       1.4\times10^{-6}]^{1/2}\ {\rm d} 
\end{eqnarray}
where $\sigma_{0,t}$ is the uncertainty in the epoch of inferior
conjunction at time $t$ in days since $T_0$.
The magnitude of $\sigma_{0,t}$ is rapidly dominated by the
$\sigma_{P_{\rm orb}}$ term, and at late times will grow linearly due
primarily to this factor, as $\sigma_{0,t}\propto n \sigma_{P_{\rm
orb}}\approx2.1\times10^{-4}\ (t/1\ {\rm yr})\ {\rm d}$.
We show the evolution of 
$\sigma_{0,t}$ (i.e. the size
of the parameter space that must be spanned by the search) as a function
of time in Figure \ref{errevolution}. 
From 2015, the effective uncertainty will grow approximately linearly with
time, doubling by 2017 and exceeding $3\times$ the current error level by
the time that full-network observations commence sometime after 2022.
Although this is less than ideal, we point out that even with the data in
hand, the ephemeris derived in this paper gives an effective uncertainty
in $T_0$ a factor of four smaller than for the previous ephemeris of SC02.

However, addtional investment in observing time will help to further
refine the orbital parameters, and improve the sensitivity of future
searches.
We have estimated the effect of additional epochs of radial velocity
measurements from 2014 onwards, based on the following assumptions.
First,
that the uncertainty on $P_{\rm orb}$ decreases as $1/\Delta T$, where
$\Delta T$ is the total span of the observations. Our results to date
indicate that this is a conservative assumption (see \S\ref{subsec2}).
Second, that the uncertainty on $T_0$ decreases as $1/n_{\rm obs}$, where
$n_{\rm obs}$ is the total number of observations. 
We consider each observing epoch to consist of 3 nights of WHT
observations, resulting in 100 radial velocity measurements, and neglect
any rejection of data required due to systematic errors around phase zero.
The resulting uncertainty curves for observations in 2014, 15, 17, 18, 21,
and 24 (roughly in between each aLIGO observing epoch) are also shown in
Fig.  \ref{errevolution}.
A relatively modest investment in observing time (6 additional epochs,
totalling 18 nights, over a decade) will allow us to maintain the $T_0$
uncertainty at or below the current level ($\approx10^{-3}$~d) throughout the
aLIGO observations. Additionally, we will incrementally improve the
uncertainty on the orbital period, down to a level of approximately
$2.3\times10^{-7}$~d.

\section{Discussion}
\label{sec2}

From analysis of spectroscopic and photometric data taken over a 12-yr
baseline, we have obtained an improved set of orbital parameters for the
X-ray binary Sco X-1.
ASAS photometric data supports the \cite{gottlieb75} orbital
period over the alias suggested by \cite{vanderlinde03} measured from the
\xte/ASM data \cite[as found by][]{hynes12}. 
Although further improvement in the orbital parameters via photometry
is unfeasible, due to large-amplitude aperiodic variations, we also
obtained an additional epoch of spectroscopic measurements which resulted
in an improvement in
the precision of the orbital period $P_{\rm orb}$ \cite[compared to that
quoted
by][]{gottlieb75} of a factor of two. We also improved
the precision of the epoch of
inferior conjunction $T_0$ (compared to that of SC02) by a factor of 2.5.

In contrast to the previous WHT/ISIS spectroscopic measurements of
SC02, which were performed with a slightly lower spectral resolution 
(0.84~\AA\ compared to 0.3~\AA\ for the 2011 WHT measurements)
we found
substantial systematic errors in the measured radial velocities close to
phase zero. We attributed these errors to the relative weakness of the
lines at this phase; the unirradiated side of the companion is facing us,
so that the line-emitting region is partially obscured.
Most notably, we found a significant variation in the measured velocity
amplitude of the Bowen lines, as well as the time of inferior
conjunction compared to the earlier ephemeris.
These variations do not seem to be related  to the systematic
uncertainties contributed to the weakness of the lines around inferior
conjunction, but instead likely indicate a change in the emission geometry
between epochs.
Variations in the Bowen line intensities on timescales as short as 1 week
offer further evidence for alterations to the emission pattern.
We also estimate that possible anisotropy of the emission with
respect to the line joining the centres of mass of the two objects, may
contribute a systematic error in the epoch of inferior conjunction of
0.005~d, or approximately 4 times the statistical uncertainty from the
spectral line fitting. It is hoped that 
additional spectroscopic data resolving the emission, together with a more
detailed model of the emission geometry,
will improve constraints on this contribution.

We also considered the impact of the system parameters on
future gravitational wave searches, as planned with
aLIGO.
\cite{watts08} quantified the search sensitivity 
via the number of model templates required to cover the
parameter space defined by the uncertainties in each of the orbital
parameters. The number of templates required for each parameter depends
linearly on the parameter uncertainty, but because correlations betwen the
parameters can be important, the number of templates for a joint subspace
is not equal to simply the product of the number of templates for each
parameter individually. That is, the number of templates for a joint
search of $P_{\rm orb}$, $T_0$ space is
\begin{equation}
N_{P_{\rm orb},T_0} \propto \Delta[P_{\rm orb}^{-2}] \Delta[T_0]
\end{equation}
where $\Delta[\lambda^i] = \lambda^i_{\rm max}-\lambda^i_{\rm min}$ for parameter
$\lambda^i$
Based on this proportionality, we can quantify the expected improvement in
sensitivity based on the fractional reduction in the number of templates
arising from the reduction in uncertainty in each parameter. This is a
factor of 
5 for the parameters listed in Table \ref{newparam}.

However, this factor only approximately represents the improvement in
sensitivity achieved for future gravitational wave searches using the
orbital parameters determined in this paper, because the effective
uncertainty in the orbital parameters (particularly $T_0$) at the epoch of
future searches must be calculated including the contributions from the
other parameters. With the results from this analysis, we expect the
improvement in the $T_0$ uncertainty from 2015 onwards, when aLIGO
observations commence, at a factor of 4 or better, giving an overall
improvement of a factor of 10 in the number of templates. Additionally, we
demonstrate in section \S\ref{future} that a modest investment in
optical observing time over the next decade can result in an improvement
likely of a factor of 
50.
We point out that this is a conservative estimate, and we expect to
improve considerably on this prediction, based on a number of additional
efforts, detailed below.

First, based on the results from our 2011 pilot observations, we will
refine our observing strategy to optimise our radial velocity
measurements. This will likely involve longer integration times for
spectra, but we will also request scheduling of our future observations to
avoid the times around phase zero.
Second, 
we plan to exploit other observing campaigns (such as the program
by which we obtained the VLT data in 2011) which can provide radial
velocity measurements with smaller uncertainties than the WHT data.
Third, we will investigate the emission line morphology via Doppler
tomography, to obtain an improved estimate of $a_x \sin i$, over that of
SC02.
Fourth, we will investigate complementary modelling efforts that can
allow us to refine our radial velocity measurements, based on inferences
of the emission pattern on the surface of the companion.
Fifth, we have an on-going program to carry out an X-ray pulsation search
of extensive archival {\it Rossi X-ray Timing Explorer}\/ data of Sco~X-1.
Improvements in the orbital ephemeris, as we have presented here, offer
improved sensitivity for X-ray pulsation searches, in an analogous manner
to gravitational wave searches. Although the pulsations and orbital
variations are effectively decoupled \cite[e.g][]{watts08}, the lack of
knowledge of the spin frequency of this source contributes the largest
share of the number of templates for the gravitational wave search. Thus,
detection of pulsations in this system, as well as being a first for a
Z-source, and a conclusive verification of the neutron-star nature of the
object, would offer the most substantial improvement in the search
sensitivity of any of the work presented here.

\acknowledgments

We are grateful to Stuart Littlefair who traded 3 hours of observing time
on the WHT to improve our coverage of the Sco X-1 orbit in 2011.
We thank the anonymous referee for their feedback, which significantly
improved this paper.
This project was supported in part by the Monash-Warwick Strategic Funding
Initiative.
DKG is the recipient of an Australian Research Council Future Fellowship
(project FT0991598).
DS acknowledges support from STFC through an Advanced Fellowship 
(PP/D005914/1) as well as grant ST/I001719/1.
JC acknowledges the support of the Spanish Ministerio de Econom\'\i{}a y
Competitividad  
(MINECO) under grant AYA2010--18080.
RC acknowledges a Ramon y Cajal fellowship (RYC-2007-01046).
This research was carried out using the 
{\sc pamela} and {\sc molly} software packages, written by Tom Marsh
(Warwick), and
also made use of the
SIMBAD database,
operated at CDS, Strasbourg, France.

Facilities: \facility{WHT}, \facility{VLT}.

\clearpage

\clearpage

\end{document}